# Revealing Gamma-Ray Binaries

I.F. Mirabel[1,2]

[1]Institut de Recherche sur les lois Fondamentales de l'Univers, Commissariat à l'Énergie Atomique et aux Énergies Alternatives, Saclay, France.
[2]Instituto de Astronomía y Física del Espacio, Conicet, Buenos Aires, Argentina.
E-mail: felix.mirabel@cea.fr

Recent ground based and space telescopes that detect high energy photons from a few up to hundreds of gigaelectron volts (GeV) have opened a new window on the universe. However, because of the relatively poor angular resolution of these telescopes, a large fraction of the thousands of sources of gamma-rays observed remains unknown. Compact astrophysical objects are among those high energy sources, and in the Milky Way there is a particular class called "Gamma-Ray Binaries". These are neutron stars or black holes orbiting around massive stars[1]. This area of high-energy astrophysics presents several challenges:
1) identifying the gamma-ray source with a source observed at other wavelengths,
2) determining the properties of the binary system, and 3) understanding the physical mechanisms by which gamma-rays are produced. In the Milky Way, only a handful of compact binaries radiating at gamma-rays have been identified (Cygnus X-3; PSR B1259-63; LSI +61° 303; LS 5039; HESS J0632+057). However, models of the evolution of massive stellar binaries suggest a much larger population of gamma-ray binaries.

The Large Area Telescope (LAT) on board of the Fermi satellite has catalogued more than 1400 high energy sources. Many of them are in the Milky Way, but because of the uncertain positions in the sky provided by the gamma-ray telescope (typically few arc-min), and the complexity of the star formation regions where Gamma-ray Binaries are usually located, the association of these high energy sources with objects observed at other wavelengths is usually uncertain. Based on the correlated orbital modulation at gamma-rays, X-rays and radio waves, the Fermi Large Area Telescope Collaboration shows[2] that the catalogued Fermi source 1FGL J1018.6-5856 is a new gamma-ray binary, demonstrating the potential of searches for periodic modulation at gamma-rays and other wavelengths to unveil new populations of gamma-ray binaries. A similar approach was successfully used to unambiguously identify with Fermi Cygnus X-3 as a source of gamma-rays[3], a microquasar[4] source of collimated relativistic jets, which was also observed at gamma-rays with the Agile satellite[5].

1FGL J1018.6-5856 is a compact object orbiting with a period of 16.6 days around a star of more than 20 solar masses. On the basis of phenomenological similarities with other gamma-ray binaries, it is most likely a pulsar that produces strong bipolar winds of particles accelerated to highly relativistic speeds by the rapidly rotating, strong magnetic field of the spinning neutron star.

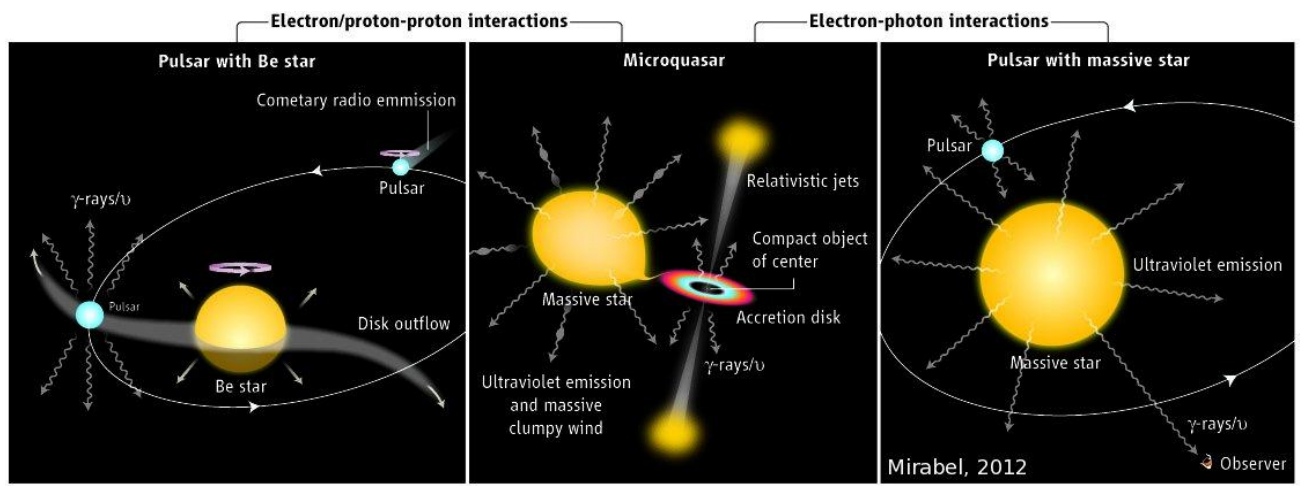

Figure 1: **Gamma-ray binaries.** Pulsar winds are powered by the rapid rotation of magnetized neutron stars. Gamma rays can be produced either by the interaction of the relativistic particles of the pulsar wind with the outflowing protons in the disk or envelope of a Be star (left panel, e.g., PSR B1259–63 and LSI +61° 303), or by their interaction with UV photons from a very massive main-sequence star (right panel, e.g. LS 5039 and 1FGL J1018.6–5856). Central panel: Microquasars are powered by compact objects (neutron stars or stellar-mass black holes) via mass accretion from a companion star. When in a microquasar the donor star is a massive star with a high-density UV flux and wind, gamma rays can be produced by electron/photon-proton and/or electron-photon interactions. Besides gamma-rays these high mass compact binaries could also be sources of high energy neutrino "ν" flux, emerging from the decays of secondary charged mesons produced at proton-proton and/or proton–gamma photon interactions.

The dominant physical mechanisms to produce the gamma-ray emission and its orbital modulation depend on the specific type of the massive star in the compact binary (see Figure 1). When the star is very massive and produces a high density field of UV photons, the main mechanism would be inverse Compton up-scattering to gamma-rays of UV photons by charged particles in the microquasar jets or pulsar winds[6,7]. In this scenario maximum of gamma-ray emission takes place when relative to the observer, the compact object is on the opposite side of the star and close to the line of sight (superior conjunction). This may occur in both types of compact binaries; in high mass microquasars as Cygnus X-3 where the compact object orbits a Wolf Rayet star, or in pulsars orbiting around very massive stars that produce high density fields of UV photons, as with the stars of type O6V((f)) in LS 5039 and 1FGL J1018.6-5856.

An alternative dominant mechanism to produce gamma rays that results in a somewhat different orbital modulation may operate when the star in the compact binary is of Be type. These stars are characterized by a massive outflow with disk and/or flattened envelope geometry, in fast rotation. Here, the gamma rays may be produced by the interaction of the pulsar wind particles with the ions in the massive outflow. This could be the case in the Be compact binaries PSR B1259–63 and LSI +61° 303, where the phasing of gamma-ray maximum at GeV energies is delayed relative to periastron[2]. Detailed hadronic mechanisms that produce gamma rays have also been proposed in a diversity of astrophysical contexts[8,9].

High-energy neutrino flux could also be produced in gamma-ray binaries of the types shown in the figure 1, emerging from the decays of secondary charged mesons produced at proton-proton and/or proton–gamma photon interactions[10]. In microquasars, relativistic protons from the jets interact with cold protons in clumps of the massive stellar wind, at large distances from the compact object [11]. In the case of a pulsar-Be binary, neutrino bursts could be produced by the interaction of relativistic protons from the pulsar wind with high-density clumps of cold protons in the massive outflowing disk or envelope of the Be star. Depending on the specific parameters of these gamma-ray binaries, it remains an open question whether neutrino signals may be detected from this type of astrophysical object.

Emission at higher energy has been detected by Cherenkov telescopes (PSR B1259-63; LSI +61° 303 and LS 5039), but it is not clear whether 1FGL J1018.6-5856 is also a TeV source. Its position is consistent with the TeV source HESS J1018-589[12], but due to possible confusion with other objects in this complex star forming region, it is not clear whether the Fermi source and a component of the HESS source are the same object. Resolving this question by using time modulation and/or more accurate positions of TeV sources, will require improving the sensitivity and angular resolution of ground based Cherenkov telescopes. The large collecting area and separation of the telescope elements in the future Cherenkov Telescope Array[13] will provide the sensitivity and angular resolution to consolidate of this emerging research area in high energy astrophysics.

**Acknowledgements:** I thank Gustavo Romero, Stéphane Corbel and Leonardo Pellizza for comments and Preston Huey for art work on the figure.